\documentclass[journal]{IEEEtran}
\usepackage{algpseudocode}
\usepackage{algorithmicx}
\usepackage{epstopdf}
\usepackage{graphics}
\usepackage{amssymb}
\usepackage{graphicx}
\usepackage{caption}
\usepackage{subfigure}
\usepackage{multirow}
\usepackage{tabularx}
\usepackage{url}
\usepackage{rotating}
\usepackage{float}
\usepackage{amsmath}
\usepackage{graphicx}
\usepackage{subfigure}
\usepackage{multirow}
\usepackage{amsfonts}
\usepackage{graphicx}
\usepackage{subfigure}
\usepackage{multirow}
\usepackage[sort, numbers]{natbib}
\usepackage{authblk}
\usepackage{multirow}
\usepackage{graphicx}
\usepackage{epstopdf}
\DeclareGraphicsExtensions{.eps}
\usepackage{longtable}
\usepackage{booktabs}

\usepackage{algorithm}
\usepackage{algpseudocode}
\usepackage{amsmath}
\usepackage{framed}
\ifCLASSINFOpdf
\else
\fi
\begin{document}
%
\title{Ancilla-Input and Garbage-Output Optimized Design of a Reversible Quantum Integer Multiplier}

\author{\IEEEauthorblockN{Jayashree HV\IEEEauthorrefmark{1}, Himanshu Thapliyal\IEEEauthorrefmark{2}, Hamid R. Arabnia\IEEEauthorrefmark{3}, V K Agrawal\IEEEauthorrefmark{4} }\\
\IEEEauthorblockA{\IEEEauthorrefmark{1}
Department of ECE, PES Institute of Technology, Bangalore, KA, India\\\IEEEauthorrefmark{2}
Department of Electrical and Computer Engineering, University of Kentucky, Lexington, KY, USA \\
\IEEEauthorrefmark{3}
Department of Computer Science, University of Georgia, Athens, GA, USA. \\
\IEEEauthorrefmark{4}
Department of ISE, PES Institute of Technology, Bangalore, KA, India
}}

\maketitle

\begin{abstract}
A reversible logic has application in quantum computing. A reversible logic design needs resources such as ancilla and garbage qubits to reconfigure circuit functions or gate functions. The removal of garbage qubits and ancilla qubits are essential in designing an efficient quantum circuit. In the literature, there are multiple designs that have been proposed for a reversible multiplication operation. A multiplication hardware is essential for the circuit design of quantum algorithms, quantum cryptanalysis, and digital signal processing (DSP) applications.  The existing designs of  reversible quantum integer multipliers suffer from redundant garbage qubits.  In this work, we propose a reversible logic based, garbage-free and ancilla qubit optimized design of a quantum integer multiplier. The proposed quantum integer multiplier utilizes a novel add and rotate methodology that is specially suitable for a reversible computing paradigm. The proposed design methodology is the modified version of a conventional shift and add method. The proposed design of the quantum integer multiplier incorporates add or no operation based on multiplier qubits and followed by a rotate right operation. The proposed design of the quantum integer multiplier produces zero garbage qubits and shows an improvement ranging from 60\% to 90\% in ancilla qubits count over the existing work on reversible quantum integer multipliers. 
\end{abstract}


%
\IEEEpeerreviewmaketitle

\section{Introduction}
A reversible logic has application in quantum computing. Reversible circuits are required to have an equal number of inputs and outputs. They are designed without any feedback and fanout. There are a few parameters or resource constraints used to measure the performance of reversible circuits namely quantum cost (QC), garbage outputs (GO), ancilla inputs (AI), gate count (GC), and delay ($ \triangle $). The quantum cost of a reversible circuit is the number of 1x1 and 2x2 quantum gates that are used to construct the circuit. Garbage outputs are the ones that are neither the primary outputs nor the ones required for further computation. An ancilla or constant inputs are required to derive a certain function and to retain one-to-one mapping. A delay corresponds to the number of primitive quantum gates in the critical path of the circuit.\\
Multipliers are the major computational units that are used frequently in DSP computations. Optimization is a major objective in designing a multiplier with design constraints. In a reversible circuit design, it is necessary to minimize the count of  garbage outputs and ancilla inputs in order to reduce the total number of qubits; therefore, we present the design of a reversible multiplier which produces zero garbage outputs and minimizes the number of ancilla inputs compared to the existing multiplier designs in the literature.\\
In this work, we present a modified version of the add and shift method of multiplication. The basic components used in our design are add or NOP block and rotate right (ROR) block. To meet our design requirement, we also present a modified circuit of  reversible ALU design presented in \cite{thomsen2010reversible}. In addition, we present a generalized circuit design methodology that is supported by a generalized behavioral model to design a constant depth rotate right reversible circuit. This design is motivated by the design presented in \cite{margolus1990parallel}. The reversible multiplier design presented in this work outperforms existing multiplier designs in terms of its garbage outputs and ancilla inputs. We also give an estimate of the gate count, quantum cost, ancilla inputs, and delay for $ N $X$ N $ qubit multiplier. In the optimization of performance parameters, there is always a trade-off; such as in optimizing one parameter, the other parameters get affected. Here, our objective is to optimize the ancilla inputs and garbage outputs, due to this the remaining parameters get affected. To indicate the trade-off, the estimation of all the performance parameters is given for each block used in designing a $ N $X$ N $ reversible multiplier.
The paper is organized into several sections namely: Section 1 provides an introduction to reversible logic gates; Section 2 elaborates on the background of reversible logic gates; Sections 3, 4, and 5 covers existing designs, behavioral model of the proposed design, and the proposed circuit design methodology, respectively; Sections 6, 7, and 8 covers performance parameters calculation, results comparison, and conclusion, respectively.
\section{Background on Reversible Logic Gates}
This section covers the basics of reversible logic gates. Any $ N $ variable reversible system is built with $ N $x$ N $ reversible circuits. A few  1x1 and 2x2 primitive quantum gates are used to construct large sized reversible gates and circuits. The quantum cost of reversible gates used in this work can be found in \cite{maslov2005reversible}. The reversible gates used in this work are Fredkin, CNOT, Toffoli, and Swap gates \cite{fredkin2002conservative, barenco1995elementary, smolin1996five}.
\subsection{CNOT Gate} 
\begin{figure}[!h]
\begin{center}
\subfigure[CNOT gate]
{\label{CNOTblock}
\includegraphics[width=1.5in]{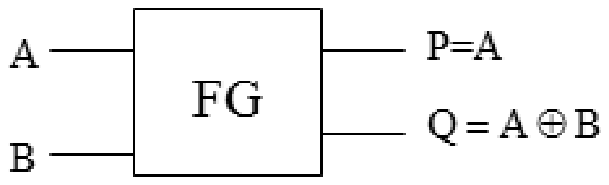}}
\quad
\subfigure[Symbol]
{\label{CNOT symbol}
\includegraphics[width=1.5in]{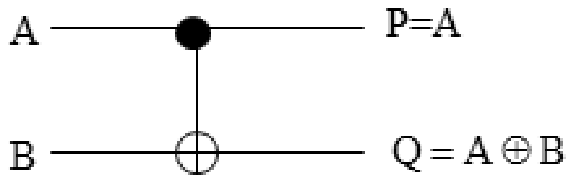}}
\end{center}
\caption{CNOT gate and its symbol}
\label{Cnotgate}
\end{figure}

A CNOT gate is also known as a Feynman gate (FG). It is a 2x2 reversible gate. The inputs and outputs are denoted as ($ A,B $) and ($ P, Q $), respectively. Here, $ A $ is treated as a control qubit, while $ B $ is treated as a target qubit. The mapping of input to outputs are denoted as  $P \leftrightarrow A $ ,  $Q \leftrightarrow A$ $ \oplus B $. The block diagram and symbol of CNOT gate are shown in Fig. \ref{Cnotgate}. QC of FG is 1.
\subsection{Toffoli Gate (TG)} 
\begin{figure}[!h]
\begin{center}
\subfigure[Toffoli gate]
{\label{tg}
\includegraphics[width=1.5in]{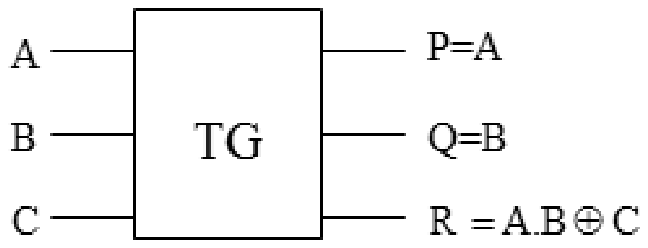}}
\quad
\subfigure[Symbol]
{\label{tgsymbol}
\includegraphics[width=0.8in]{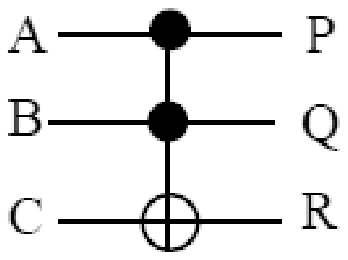}}
\end{center}
\caption{Toffoli gate and its symbol}
\label{tggate}
\end{figure}

This gate is also known as a $C^{2} $NOT gate. The TG used in our work is a 3x3 gate with  inputs ($ A, B, C $) and outputs ($ P, Q, R $), respectively. Here, $ A$ and $B $ are the control qubits, while $ C $ is the target qubit. The mapping between inputs and outputs is given with the relation $P  \leftrightarrow  A$,  $Q \leftrightarrow B $,  $ R \leftrightarrow $ $(A \cdot B) \oplus C$. The block diagram and symbol of TG are presented in Fig. \ref{tggate}. QC of TG is 5.
\subsection{Fredkin Gate (FRG)} 
\begin{figure}[!h]
\begin{center}
\subfigure[Fredkin gate]
{\label{frg}
\includegraphics[width=1.5in]{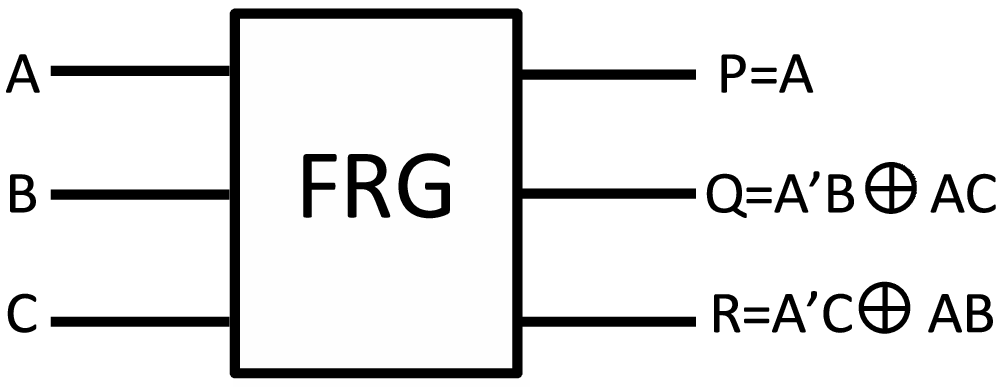}}
\quad
\subfigure[Symbol]
{\label{frggsymbol}
\includegraphics[width=1in]{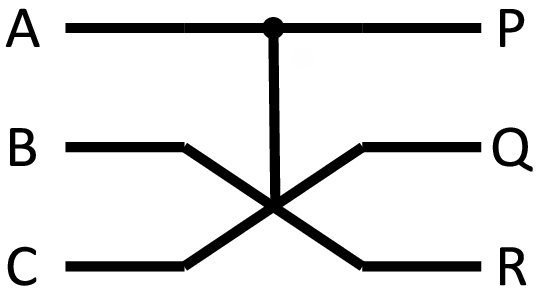}}
\end{center}
\caption{Fredkin gate and its symbol}
\label{figfredkingate}
\end{figure}

A Fredkin gate is commonly used as a controlled Swap gate. In this paper, we use a 3x3 Fredkin gate. $ A, B$, and $C $ are the inputs and $ P, Q,$ and $R $ are the output qubits. The mapping of the input and outputs are given based on the value of $ A $, which is the control qubit. When $  A $ is high, $ Q \leftrightarrow C $ and $ R \leftrightarrow B $. When $ A $ is low, $ Q \leftrightarrow B $ and $  R \leftrightarrow C $. Irrespective of $ A $ value, $ P \leftrightarrow A $. The block diagram  and symbol of FRG gate are shown in Fig. \ref{figfredkingate}. QC of FRG is 5.
\subsection{Swap Gate (SG)} 
\begin{figure}[!h]
\begin{center}
\subfigure[Swap gate]
{\label{sg}
\includegraphics[width=1in]{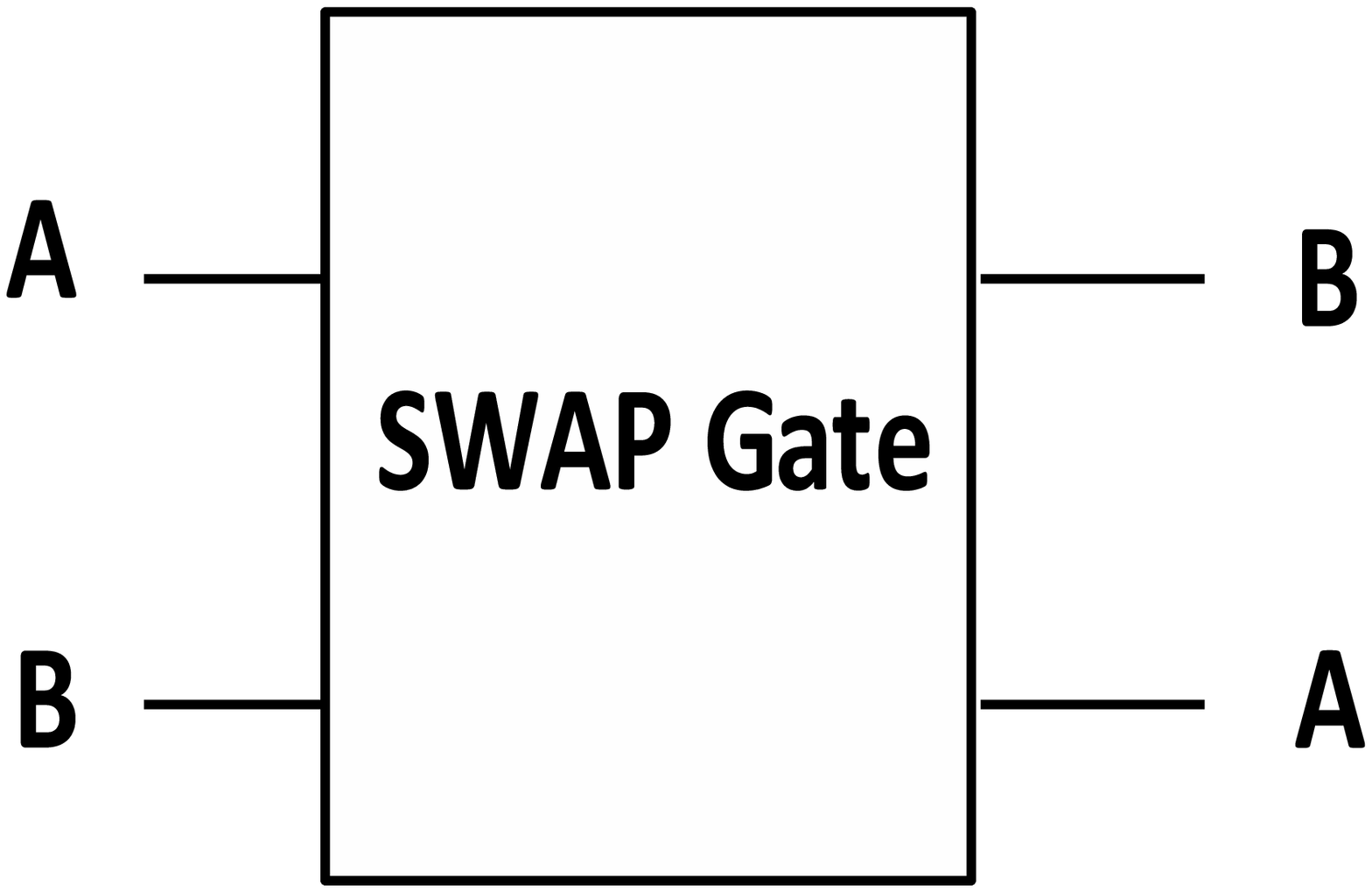}}
\quad
\subfigure[Symbol]
{\label{sgsymbol}
\includegraphics[width=0.8in]{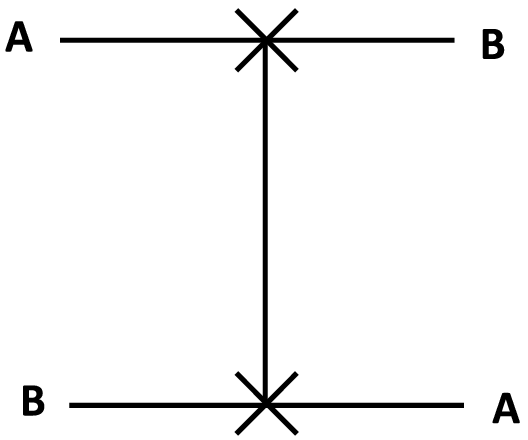}}
\end{center}
\caption{Swap gate and its symbol}
\label{figswapgate}
\end{figure}

The Swap gate is a 2x2 gate. It swaps the input and output qubits unconditionally. The mapping is $ A \leftrightarrow Q $ and $ B \leftrightarrow P $. The block diagram and symbol are shown in Fig. \ref{figswapgate}. QC of SG is 3.
\section{Existing Work}
The research on reversible logic is being explored in the domains of design, synthesis, and testing. Although there are many synthesis techniques available to realize reversible circuits, having dedicated designs of a reversible circuit component gives flexibility in choosing the designs based on the application requirement.  Multiple ways of designing arithmetic circuits have been explored in conventional, reversible and quantum computing \cite{draper2006logarithmic,takahashi2005linear,takahashi2009quantum, takahashi2009quantum1,choi2011effect,vedral1996quantum,thapliyal2013design,thapliyal2013progress,thapliyal2006combined,thapliyal2006reduced,
thapliyal2006reversible,thapliyal2005reversible,ParallelAdderThapliyal,thapliyal2005reversiblemultiplier,thapliyal2009efficient}.  
Several interesting contributions have been explored in the existing synthesis of reversible logic circuits \cite{Markov,golubitsky2012study,maslov2004reversible,maslov2005reversible,yang2008bi,maslov2011reversible,maslov2015advantages, Saeedi_2010,HungTCAD06}.  
In this section, we discuss the existing reversible multiplier designs that are important arithmetic circuits in processing digital signals. \\
There are several multiplier designs proposed by many authors in view of optimizing different performance parameters namely quantum cost, ancilla inputs, garbage outputs, logic depth, and a combination of these parameters. The multipliers proposed in \cite{zomorodi2012ultra, bhagyalakshmi2012optimized, panchal2014analysis, rangaraju2013design,mamataj2015approach} follow two phases in computing product terms. In the first phase, the partial products are computed. In the second phase, the summation of partial products are computed to get the final product terms. In all the designs mentioned above, the partial product generation and summation stages are improved either in terms of constant inputs, quantum cost, or garbage outputs. We found that the design in  \cite{mamataj2015approach} gives better results when compared to other existing work  in terms of ancilla inputs and gate count. Apart from the regular parallel or array multiplier designs, other techniques of multiplications like Booth, Wallace, and Vedic are proposed by researchers in \cite{sultana2015analysis, thapliyal2006novel, banerjee2013design}. All these designs are illustrated for smaller operand width. It is necessary for any designer to choose designs based on their performance over a wide range of operand width. We found that the recent publication on multiplier in \cite{kotiyal2014circuit} discussed $ N$x$N $ reversible multiplier design. We considered the designs proposed in  \cite{kotiyal2014circuit, zhou2011transistor} and compared it with our work, as the efforts in both of these papers were to reduce the number of constant inputs and garbage outputs. 
\section{Proposed Reversible Multiplier Behavioral model}
\label{secmulbehavior}
In this section, we present an algorithm for multiplying two $ n $ qubit numbers $ A $ and $ B $. The result is stored in $ 2n $ qubit product register $ P $. There are two conventional techniques of add and shift method of multiplying two numbers. Shift the multiplicand left and add it to the product register contents iteratively  or add the multiplicand and shift the product register contents right iteratively. In the first technique, after the computation is completed, the multiplicand will not be in its original form and since one cannot recover the multiplicand, it will be considered as garbage qubits.  We choose the second technique. The final result will be the contents of the product register and the multiplicand contents are unaltered, so that garbage outputs are not generated.
\subsection{Behavioral Model of NxN  Multiplier}
\begin{figure}[!h]
\begin{center}
\includegraphics[width=1.8in]{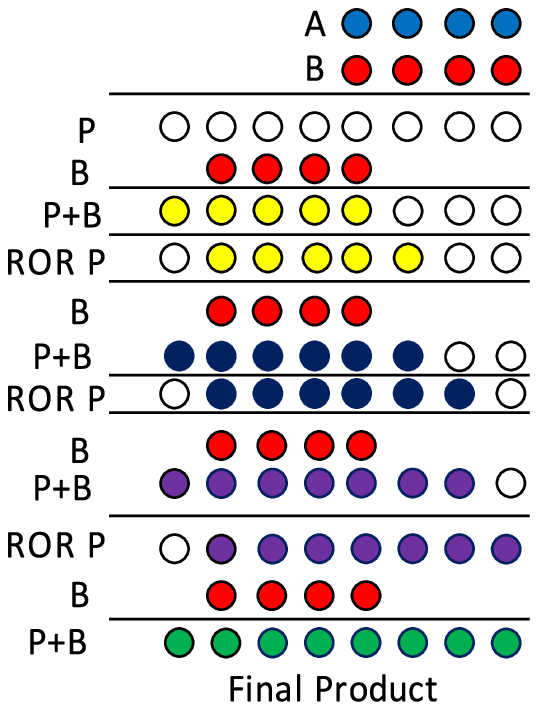}
\end{center}
\caption{4x4 qubit multiplication dot diagram}
\label{figmuldot}
\end{figure}

\begin{algorithm}
\caption{Add and rotate method to model $ N$x$N $ Multiplier}
\label{multiplieralgo}
\begin{algorithmic}
\Function{Multiplier}{$\left| A_{n} \right\rangle $, $\left| B_{n} \right\rangle $, $\left| P_{2n} \right\rangle $=$\left| 0_{2n} \right\rangle $} 

\For{$i=0$ to $n-2$}
\If{$\left|A_{[i]}\right\rangle=\left|1\right\rangle$ }
\State {$\left| P_{[2n-1:n-1]}\right\rangle $ $=$ $\left| P_{[2n-1:n-1]} \right\rangle+ \left| B_{[n-1:0]} \right\rangle $};
\EndIf 
\State{$\left| P_{[2n-1:0]}\right\rangle $= \Call{Rotateright}{$\left| P_{[2n-1:0]}\right\rangle $}};
\EndFor
\If{$\left|A_{[n-1]}\right\rangle=\left|1\right\rangle$ }
\State {$\left| P_{[2n-1:n-1]}\right\rangle $ $=$ $\left| P_{[2n-1:n-1]} \right\rangle+ \left| B_{[n-1:0]} \right\rangle $};
\EndIf\\
\Return $ P $;
\EndFunction
\end{algorithmic}
\end{algorithm}
The multiplier algorithm is best explained using an example of 4x4 multiplication with a dot diagram as shown in Fig. \ref{figmuldot}. Initially, the $ P $ register is loaded with ancilla 0 qubits. The multiplicand $ B $ is added to the $ P $ register contents. For the $ P $ register, it considers a least significant position (LSP) from $ n-1 $ and move up to $ 2n-1 $ position. For the $ B $ register, LSP starts from 0 and moves up to $  n-1 $. A multiplicand is added to the $ P $ register only if the corresponding multiplier qubit is high; otherwise, only rotate operation is performed. The rotate right operation is performed irrespective of the value of a multiplier qubit. While adding a multiplicand to the $ P $ register, the $ \left( n-1 \right) ^{th} $ position of the $ P $ register is aligned to the $ 0^{th} $ position of the $ B $ register. Due to this alignment, one rotate operation is eliminated at the end of computation. The diagram shown in Fig. \ref{figmuldot} is self-explanatory of the algorithm.

\subsection{Behavioral Model of Rotate Right operation}
\begin{figure}[!h]
\begin{center}
\includegraphics[width=3.5in]{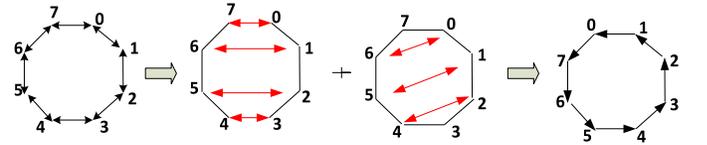}
\end{center}
\caption{Rotate right with two sets of disjoint transpositions}
\label{figshifterpermutation}
\end{figure}

In this section, we present a rotate right operation for 2$ n $ qubit data width. In the multiplication technique presented in the Algorithm \ref{multiplieralgo}, there is a need to rotate the $ P $ register contents to the right; the size of the $ P $ register is 2$ n $ qubit width. The rotate right operation is performed by swapping the qubits in two stages. The circuit is designed to obtain  constant logic depth. To give an illustrative example, we present a rotate  right operation in Fig. \ref{figshifterpermutation} for  data width of 8 qubits. The numbers 0 to 7 represent the qubit positions. The initial representation of qubits is shown in the left most part of Fig. \ref{figshifterpermutation}. The rotate right operation is performed in two stages. In the first stage, qubits in the position pairs (0,7), (1,6), (5,2), and (4,3) are swapped [(0,7) indicates 0 is swapped with 7]. In the second stage, (0,6), (1,5), and (2,4) are swapped. It is visible from the diagram that the swapping of qubits in each stage is parallel which reduces the logic depth compared to the sequential shifting of qubits. The reversible circuit design of rotate right operation will be discussed in the latter section of this paper. We present a generalized pseudo code for the method discussed in  Fig. \ref{figshifterpermutation} to swap the qubits in two stages.
The pseudo code presented in the Algorithm \ref{rotatealgo} performs rotate right operation by 1 qubit position. This code works for both even and odd data width. The Algorithm \ref{rotatealgo} will be useful in any application when data input width is variable (i.e. even or odd). For the multiplication technique proposed in this paper, the width of product register ($ P $) is always even; hence, the second half of the pseudo code is redundant for the proposed work.
\begin{algorithm}[h]
\caption{Pseudo Code for Rotate Right Operation}
\label{rotatealgo}
\begin{algorithmic}
\State \Call {ROTATERIGHT}   {$ \left| P \right\rangle   $}
\State $ k $=\Call {Sizeof}{$\left|P\right\rangle $};\Comment{$ k  $ is an integer}
\State $k1$=\Call{floor}{$ k $/2}; \Comment{$ k1 $ is an integer}
\If{$ k $ mod $ 2 $ == $0 $}     \Comment{For even number of qubits}
\State $ i=0 ;$  $ j $=$ k-1; $  \Comment{$ i $ and $ j $ are integers} 
\While{$i<k1$ $\&\&$ $j>=k1$ } \Comment{First Stage}
\State \Call{Swap}{$\left| P_{[i]}\right\rangle $,$\left| P_{[j]}\right\rangle $};
\State $i=i+1;$  $j=j-1;$
\EndWhile
\State $ i=0 ;$ $ j=k-2 $;
\While{ $i<k1-1$  $\&\&$  $j>=k1$   } \Comment{Second stage}
\State \Call{Swap}{$\left| P_{[i]}\right\rangle $$\left| P_{[j]}\right\rangle $};
\State $ i=i+1; $ $ j=j-1; $
\EndWhile
\Else  \Comment{For odd number of qubits}

\State $i=0;$ $j=k-1;$
\While {$i<k1$  $ \&\& $  $j>=k1+1$} \Comment{First Stage}
\State \Call{Swap}{$\left| P_{[i]}\right\rangle $,$\left| P_{[j]}\right\rangle $}
\State $i=i+1;$ $j=j-1$
\EndWhile
\State $ i=0;$ $ j=k-2;$
\While {$i<k1 $ $ \& \& $ $ j>=k1 $}   \Comment{Second Stage}
\State \Call{Swap}{$\left| P_{[i]}\right\rangle $,$\left| P_{[j]}\right\rangle $}
\State $ i=i+1 ;$ $ j=j-1;$
\EndWhile
\EndIf\\
\Return $ P $;
\end{algorithmic}
\end{algorithm}
%
%
%
%
%
%
%
%
%

\section{Proposed Garbageless Reversible Multiplier Circuit Design}
From the behavior models Algorithm \ref{multiplieralgo} and Algorithm \ref{rotatealgo} presented in the Section \ref{secmulbehavior}, it is clear that to compute the product of two n qubit numbers, we need  to design the following reversible circuits: (1) $ n $ qubit addition or no operation (ADD/NOP) circuit (2) uncontrolled rotate right operation circuit. This section elaborates on the reversible circuit design methodology of ADD/NOP and rotate right (ROR) block.

\subsection{ADD or NOP Circuit Design}\label{ADD/NOP}
\begin{figure}[!h]
\begin{center}
\includegraphics[width=3.6in]{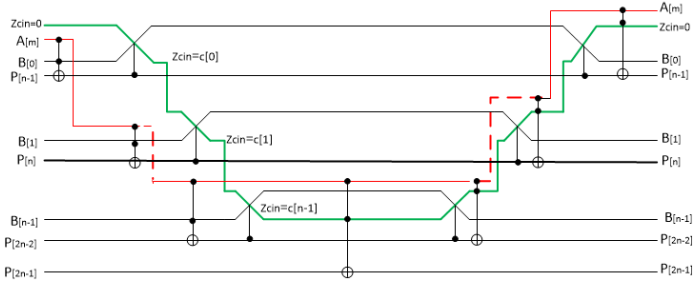}
\end{center}
\caption{Reversible ADD/NOP circuit}
\label{figaddnop}
\end{figure}

ADD or NOP block has evolved from the ALU design proposed in \cite{thomsen2010reversible}. We have modified the original work to adapt to our garbageless multiplier design. The reversible circuit design of ADD/NOP block is shown in Fig. \ref{figaddnop}. 
Inputs to the ADD/NOP block are:\\
 (a) $ n $ qubit product register $\left| P_{[2n-1:n-1]}\right\rangle $, (b) $ n $ qubit input operand $\left| B_{[n-1:0]}\right\rangle $, (c) 1 qubit input $ Zcin $ initialized with ancilla 0 , (d) 1 qubit input operand $ A_{[m]} $, where $ m $ is the qubit position varying from 0 to $ n-1 $.
Here, $ A_{[m]} $ acts as the control qubit; if it is  high, the $ P $ and $ B $ register contents are added. At the output, we have the $ B $ register contents unaltered, where the $ P $ register contents will have the sum of $ P $ and $ B $ contents. If $ A_{[m]} $ is low, then the  $ B $ and $ P $ register contents are regenerated at the output without modification. The role of $ Zcin $ is to propagate the carry generated from the previous qubit position. If the control qubit $ A_{[m]} $ is high, $ P_{[2n-1]} $ will have the final carry out generated; otherwise, it will retain its value.

The computation of ADD/NOP block is summarized below. Here, the index of B varies from 0 to $ n-1 $ and $ P $ varies from $ n-1 $ to $ 2n-1 $, according to the requirement of the multiplier design. The index of $ A $ is chosen to be $ m $ which ranges from 0 to $ n-1 $, where $ n $ is the size of operands (multiplier and multiplicand).  Here, $ j $ is used to indicate the qubit position of the product register $ P $.
\begin{enumerate}
\item \textbf{Computation Phase 1}:\\ Initialize $ Zcin $ with ancilla $ 0 $ qubit. In further stages, the same line will propagate the  carry generated from the previous stage.
\item Step 1:  Apply 3x3 Toffoli gate at locations  $  A_{[m]} $, $ B_{[0]} $, and $ P_{[n-1]} $. After the computation, $  A_{[m]} $ and $  B_{[0]} $ will retain their  value. $ P_{[n-1]} $ will get transformed  according to the equation given below.
\begin{eqnarray}
\left | P_{[n-1]} \right \rangle =\left(\left |    A_{[m]}  \right \rangle \cdot \left |   B_{[0]}  \right \rangle \right) \oplus \left | P_{[n-1]} \right \rangle
\end{eqnarray}

\item Step 2a: For $ 0 $ $ \leq $ $ i $ $ \leq $ $ n-1 $  and $ n-1 $ $ \leq $ $  j $ $ \leq $ $ 2n-2 $, apply 3x3 Fredkin gate at locations $ P_{[j]} $, $ B_{[i]} $, and $ Zcin $. Here, $ P_{[j]} $ acts as a control line to FRG gate and it will not change after the computation. The remaining lines $ B_{[i]} $ and $ Zcin $ will get modified according to the equations shown below. 
\begin{equation}
    \left | Zcin \right \rangle = 
\begin{cases} 
 \left |B_{[i]}\right \rangle  & \mbox{if  $ \left | P_{[j]} \right \rangle $ = 1 } \\
\left | Zcin \right \rangle & \mbox{if $ \left | P_{[j]} \right \rangle $ = 0 } 
\end{cases}
\end{equation}
\begin{equation}
    \left | B_{[i]} \right \rangle = 
\begin{cases} 
 \left | Zcin \right \rangle  & \mbox{if  $ \left | P_{[j]} \right \rangle $ = 1 } \\
\left | B_{[i]} \right \rangle & \mbox{if $ \left | P_{[j]} \right \rangle $ = 0 } 
\end{cases}
\end{equation}
\item Step 2b: For $ 1 $ $ \leq $ i $ \leq $ $ n-1 $  and $ n $ $ \leq $ j$ \leq $ $ 2n-2 $, apply a 3x3 Toffoli gate at locations $  A_{[m]} $, $ B_{[i]} $, and $ P_{[j]} $. After the computation, $  A_{[m]} $ and $ B_{[i]} $ will retain their  value. $ P_{[j]} $ will get transformed  according to the equations given below.
\begin{equation}
   \left | P_{[j]}\right \rangle = 
\begin{cases} 
 \left | P_{[j]}\right \rangle \oplus   \left | B_{[i]} \right \rangle & \mbox{if  $ \left |  A_{[m]} \right \rangle $ = 1 } \\
 \left | P_{[j]} \right \rangle & \mbox{if $ \left |  A_{[m]} \right \rangle $ = 0 } 
\end{cases}
\end{equation}
\textit{Step 2a and Step 2b execute concurrently}.
\item Step 3: Apply a 3x3 Toffoli gate at locations $  A_{[m]} $, $ B_{[n-1]} $, and $ P_{[2n-1]} $ . This step is required in order to store the final carry out after $ n $ qubit addition. After the computation, $ P_{[2n-1]} $ stores final carry out.
\begin{equation}
    \left |P_{[2n-1]}\right \rangle = 
\begin{cases} 
  \left | B_{[n-1]} \right \rangle  & \mbox{if  $ \left |  A_{[m]} \right \rangle $ = 1 } \\
 \left |P_{[2n-1]} \right \rangle & \mbox{if $  \left |  A_{[m]} \right \rangle $ = 0 } 
\end{cases}
\end{equation}
In other words, $ P_{[2n-1]} $ stores the final carry out only if the control line that is corresponding to the  multiplier qubit $  A_{[m]} $ is high; otherwise, it will restore the previous value present in it, that  means, it will retain the previous carry value stored in $ P_{[2n-1]} $ from the previous computation.
\item \textbf{Computation Phase 2}:\\The steps in this phase contribute to the generation of the final product value $ P $ and the  regeneration of the contents of multiplicand $ B $ . 
\item Step 4a: For $ 2n-2 \ge j \ge n-1 $ and $ n-1 \ge i \ge 0 $, apply a 3x3 Fredkin gate at locations $ P_{[j]} $ ,  $ B_{[i]} $, and $ Zcin $. After the computation, the value at $ P_{[j]} $ will be retained as it is where as $ Zcin $ and $ B_{[i]} $ will get modified according to the  equations shown in Computation Phase 1 Step 2a.
\item Step 4b: For $ 2n-2 \ge j \ge n-1 $ and $ n-1 \ge i \ge 0 $, apply a 3x3 Toffoli gate at locations $  A_{[m]} $, $ Zcin $, and $ P_{[j]} $. At the end of computation, $  A_{[m]} $ and $ Zcin $  will retain its value; where as, $ P_{[j]} $ will get modified according to the equation mentioned below.
\begin{equation}
    \left |P_{[j]} \right \rangle = 
\begin{cases} 
  \left |Zcin\right \rangle \oplus \left |P_{[j]}\right \rangle  & \mbox{if  $ \left |   A_{[m]} \right \rangle $ = 1 } \\
 \left |P_{[j]}\right \rangle & \mbox{if $ \left |  A_{[m]} \right \rangle $ = 0 } 
\end{cases}
\end{equation}
\end{enumerate}
\textit{Steps 4a and 4b execute sequentially}.

\subsection{Rotate Right Reversible Circuit Design}
\begin{figure}[!h]
\begin{center}
\includegraphics[width=1.8in]{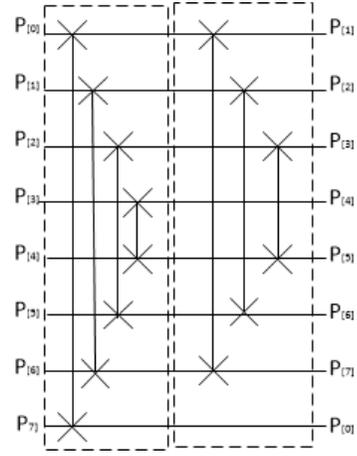}
\end{center}
\caption{Reversible rotate right circuit(ROR)}
\label{figshifter1}
\end{figure}

The reversible circuit for the rotate right operation is shown in Fig. \ref{figshifter1}. The circuit takes no ancilla and rotates the data to the right from MSB qubits to LSB qubits by 1 position (ROR). The reversible rotate circuit is designed using Swap gates and performs a rotate operation with constant delay.
For clarity of understanding, we have shown 8 qubit ROR circuit design. The product register qubits $ P_{[0]} $ to $ P_{[7]} $ are given as input. After one rotate operation, $ P_{[0]} $ occupies $ {P_{[7]}}^{th}$ qubit position and qubits from $ P_{[7]}$ to $ P_{[1]} $ shift to the right by one position. The quantum cost of Swap gate is 3. The delay in performing the rotation operation involves two Swap gates in series; therefore, the constant delay of 6 is obtained by considering each cycle individually and decomposing it into two sets of disjoint swaps. The gates shown in the dotted boxes are executed in parallel. The rotator design is motivated by the property proven in \cite{margolus1990parallel}. According to the authors, any permutation is the composition of two set of disjoint transpositions. This is illustrated in Fig. \ref{figshifterpermutation}, which also shows that the cycle is a composition of two reflections. The authors have proven that any  permutation of $ n $ qubits can be performed in 4 layers (levels or logic depth) of CNOT gates with $ n $ ancilla input qubits, or in 6 layers with no ancilla input qubits (delay of 2 Swap gates). If we had opted for first technique of multiplication in which the multiplicand ($ n $ bit operand) is shifted, it leaves the multiplicand altered, which in turn will yield garbage or garbage output.

\begin{figure}[!h]
\begin{center}
\includegraphics[width=2in]{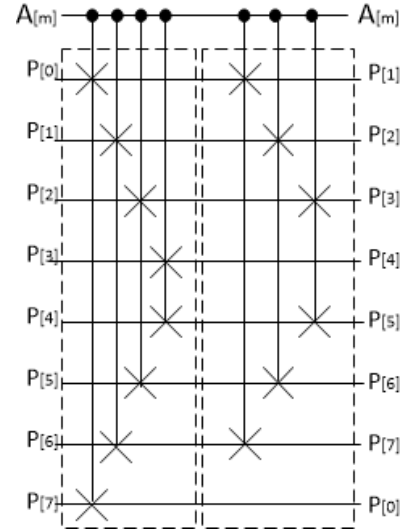}
\end{center}
\caption{ Alternative reversible rotate circuit}
\label{figshifter2}
\end{figure}

The rotate circuit proposed performs an unconditional rotate operation in the sense that irrespective of multiplier qubit value, a rotation is performed. Another option that we explored was to use a conventional multiplication technique that needs conditional shift or rotate circuit. A controlled Swap gate or Fredkin gate instead of Swap gate can be used to rotate the qubits. The rotate operation is controlled by $  A_{[m]}$ qubit. The same control qubit is used by all the Fredkin gates in the rotate circuit as shown in Fig. \ref{figshifter2}. Here, the computation becomes sequential and the delay will increase with the size of the rotate circuit unlike our proposed design. Another reason for ignoring the design shown in Fig. \ref{figshifter2} is that the quantum cost of Fredkin gate is more than the Swap gate. To optimize the delay and quantum cost, we omitted this option. If the delay has to be maintained constant, then one has to store the control lines for these Fredkin gates and use them in parallel. This will again increase the number of ancilla lines, thus violating our objective of minimizing the ancilla lines.
\subsection{Reversible Multiplier Circuit Design Methodology}
In this section, we illustrate the design steps of a reversible multiplier. 
\begin{figure}[!h]
\begin{center}
\includegraphics[width=3.5in]{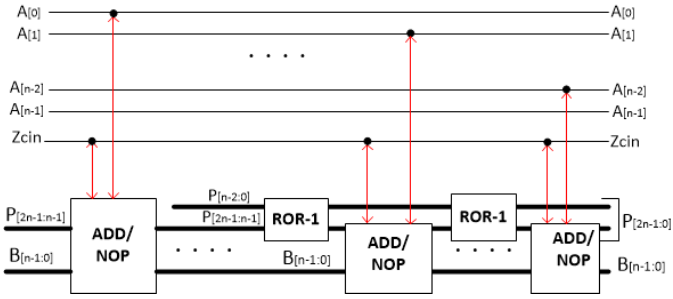}
\end{center}
\caption{Reversible multiplier circuit}
\label{figmultiplier}
\end{figure}

\begin{enumerate}
\item For $ m=0 $ to $ n-2 $  repeat Step-1 and Step-2
\item Step 1:\textbf{ADD or NOP} \\
Apply the data qubits $   A_{[m]} $, $ Zcin $, product register $ P_{[2n-1:n-1]}$, and the multiplicand register contents $ B_{[n-1:0]} $ to ADD/NOP block. After the computation, the contents of  $  A_{[m]} $, $ Zcin $, and $ B $ are restored; where as, the $ P $ register contents will get modified according to the  computation equations mentioned in the Section \ref{ADD/NOP}. ADD/NOP circuit will perform the addition on $ P $ and $ B $ register contents if $  A_{[m]} $= high; otherwise, the contents of those registers are retained. 

\item Step 2: \textbf{Rotate Right (ROR)}\\ Apply the P register contents to ROR (ROR-1 block indicates rotate right by 1 position) block, which performs a rotate right operation with a constant delay of 6. The computation is carried out as follows:\\
$\left| P_{[2n-1:0]}\right\rangle $ $ \leftarrow $ $\left| P_{[2n-1:0]}\right\rangle$ $\circlearrowright$1;
\item Step 3: update $ m=n-1 $, repeat Step-1.
\end{enumerate}
\section{Performance Parameters calculation}
In this section, we discuss the performance parameters and the calculation for each circuit used in the reversible multiplier design. As a final part of the calculation, we show the overall calculation of the reversible multiplier.
\subsection{Performance Parameters of ADD/NOP Block}
\begin{figure}[!h]
\begin{center}
\includegraphics[width=3.6in]{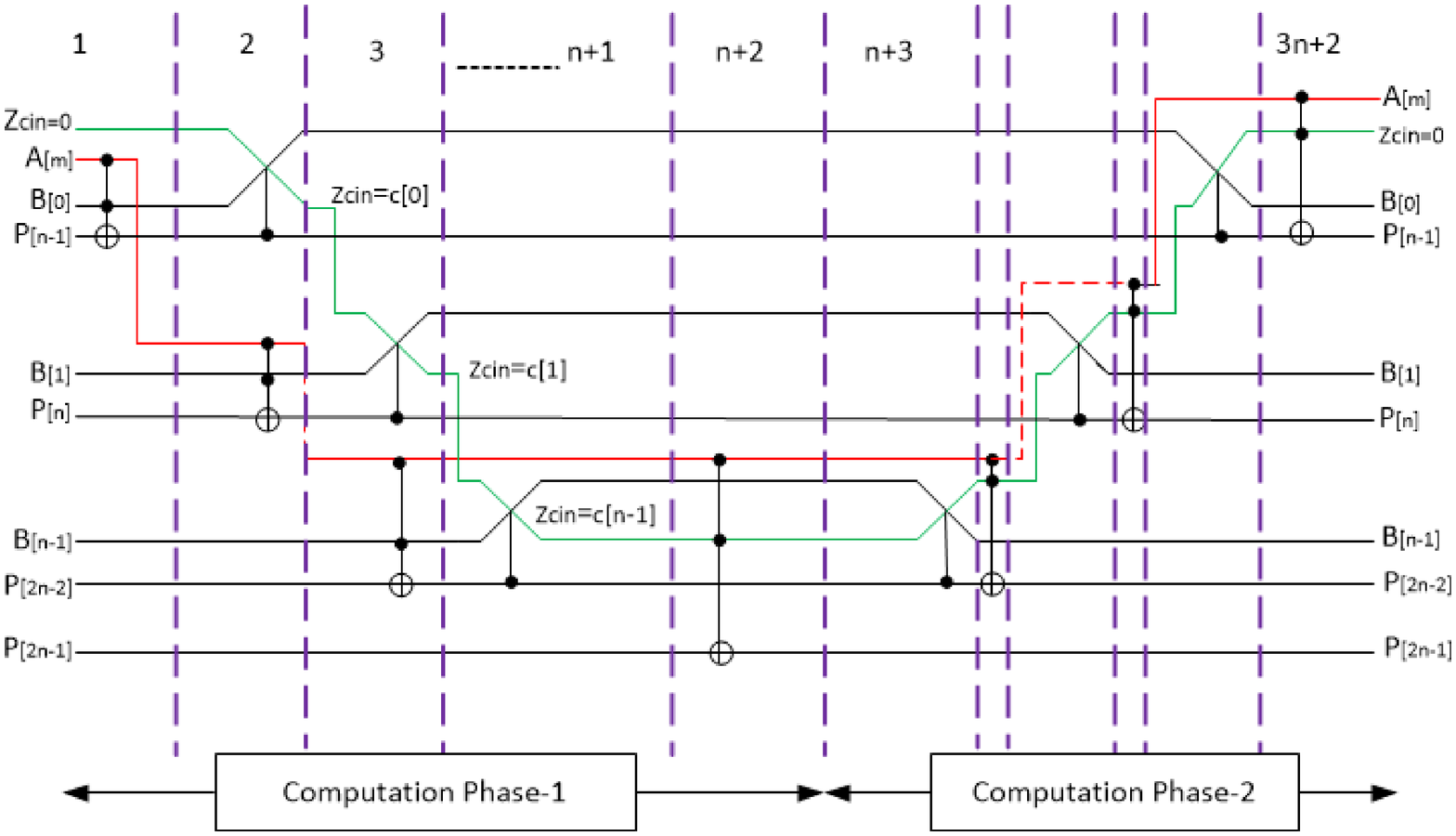}
\end{center}
\caption{Critical Path Computation}
\label{figaddnopcriticalpath}
\end{figure}

The equations shown below are with respect to the design mentioned in Fig. \ref{figaddnop}. The calculation of quantum cost (QC) is shown below.
\begin{align} 
\mbox {QC(ADD/NOP)} &=5 \ast \mbox{No. of TG} +5\ast \mbox{No. of FRG}\nonumber\\ 
 &=5\ast(2n+1)+5\ast(2n)\nonumber\\
 &=20n+5
 \end{align}
The ancilla inputs include the product register qubits ($ n $+1) which are initially set to ancilla 0 and $ Zcin $ used for carry propagation initialized to ancilla 0. So the ancilla for $ n $ qubit ADD/NOP block is given below.
\begin{align} 
\mbox {AI(ADD/NOP)} = n+2 
  \end{align}
The delay of ADD/NOP block includes the critical path delay. To find the critical path, the design has been divided into stages where each stage is sequential in execution. This is illustrated in Fig. \ref{figaddnopcriticalpath} in vertical lines. The computation stages are divided into Phase 1 and Phase 2. The Phase 1 consists of the computation of half sum and final carry out. In Phase 2, the full sum is computed and the content of the $ B $ register  is regenerated. There are $ 3n+2 $ stages. We have computed the  number of stages involving Toffoli gates and Fredkin gates. Since the delay is proportional to the quantum gates present in each reversible gate, the total delay of ADD/NOP circuit can be found by using the equation shown below.
\begin{align} 
\mbox {$ \bigtriangleup $(ADD/NOP)} &=\mbox{Delay of single stage} \ast \mbox{No. of stages} \nonumber\\ 
 &=5\ast(3n+2)\nonumber\\
 &=15n+10
 \end{align}
  
%
\subsection{Performance Parameter of ROR Block}
The performance parameters are estimated for the rotate right reversible circuit (ROR). It is discussed in the previous sections that to perform one-time rotation, two phases of swap operations are carried out. The swapping of qubits in each phase are computed parallely. The rotate circuit presented has no garbage outputs and no ancilla input qubits. Qcost (QC) and delay ($ \bigtriangleup $) are calculated as follows.
\begin{align} 
\mbox {QC(ROR)} &=3\ast \mbox{No. of SG}\nonumber \\ 
 &=3\ast(n-1)\nonumber\\
 &=3n-3
  \end{align}
 The above equation is for a generalized rotate circuit design. For our work, we feed the $ 2n $ qubits of the product register contents to the rotate block. The modified equation is shown below.
\begin{align} 
\mbox {QC(ROR)} &=3\ast \mbox{No. of SG}\nonumber \\ 
 &=3\ast(2n-1)\nonumber\\
 &=6n-3
  \end{align}
  \begin{align} 
  \mbox {$ \bigtriangleup $(ROR)} =6 
      \end{align}
\subsection{Performance Parameters of NxN Reversible Multiplier Block}
The performance parameters calculation for $ N $x$ N $ reversible multiplier block is generated by summing up the calculations of ADD/NOP and Rotate Right block (ROR) components.
\begin{align} 
\mbox {QC(Mul)} &=n\ast \mbox{QC(ADD/NOP)}+(n-1)\ast \mbox{QC(ROR)}\nonumber \\ 
 &=n\ast(20n+5)+(n-1)\ast(6n-3)\nonumber\\
&=26n^{2}-4n+3
 \end{align}
 Although the ancilla inputs of the rotate right circuit is nil, the input to the ADD/NOP block and ROR circuit is the $ P $ register contents, and all the $ 2n $ qubit locations of $ P $ register which are initialized with ancilla 0 qubits.
 \begin{align} 
 \mbox {AI(Mul)} &=\mbox{AI(ADD/NOP)}+ \mbox{AI(ROR)} \nonumber\\ 
  &=2n+1
  \end{align}
   The delay of the multiplier is the summation of delay of ADD/NOP block and rotate right circuit.
  \begin{align} 
    \mbox {$ \bigtriangleup $(Mul)} &=n \ast \mbox{$ \bigtriangleup $(ADD/NOP)}+(n-1)\ast \mbox{$ \bigtriangleup $(ROR)}\nonumber \\ 
    &=n \ast (15n+10)+(n-1)\ast 6\nonumber\\
    &=15n^{2}+16n-6
    \end{align}
\section{Comparison Results}
In the literature, there are more designs presented for a 4x4 reversible multiplier. It is necessary to design  a circuit which is scalable to any size. Hence, we compared our design with other $ N $x$ N $ reversible designs that are available in the literature. The designs proposed in \cite{kotiyal2014circuit} and \cite{zhou2011transistor} show the calculation for $ N $x$ N $ reversible multiplier. In both of these papers, the authors have shown only the ancilla inputs and garbage outputs calculations; the comparisons shown in Tables \ref{tabmulcomp}, \ref{tabmulcomp2} list only ancilla inputs and garbage outputs for these papers. It is clear from the result shown in Table \ref{tabmulcomp} that as the operands size increases, the percentage of improvement also increases. Our proposed design showed a better improvement in terms of the ancilla inputs resulting in saving the chip area since the number of lines are reduced.\\
We have listed the garbage outputs of the designs proposed in \cite{kotiyal2014circuit} and \cite{zhou2011transistor}. Our design outperforms the existing designs because it is 100\% better in terms of garbage outputs since our design produces no garbage.
We compared our design with another garbageless reversible multiplier design using the recursive scheme in \cite{portugal2006reversible}.
\begin{table}[h]
\centering
\caption{Ancilla inputs comparison of  $ N$x$ N $ Reversible Multiplier}
\label{tabmulcomp}
\begin{tabular}{|c|c|c|c|c|c|}
\hline
\begin{tabular}[c]{@{}l@{}}$ N $ \end{tabular} & \begin{tabular}[c]{@{}l@{}}Ancilla\\    inputs\\in    {[}1{]}\end{tabular} & \begin{tabular}[c]{@{}l@{}}Ancilla \\     inputs\\  in {[}2{]}\end{tabular} & \begin{tabular}[c]{@{}l@{}}Ancilla \\     inputs\\   in {[}3{]}\end{tabular} & \begin{tabular}[c]{@{}l@{}}\%imp\\      over\\       {[}2{]}\end{tabular} & \begin{tabular}[c]{@{}l@{}}\%imp\\      over\\       {[}3{]}\end{tabular} \\ \hline
4 & 9 & 23 & 28 & 60.86 & 67.85 \\ 
8 & 17 & 83 & 120 & 79.51 & 85.83 \\ 
16 & 33 & 303 & 496 & 89.10 & 93.34 \\ 
32 & 65 & 1135 & 2016 & 94.27 & 96.77 \\ 
64 & 129 & 4351 & 8128 & 97.03 & 98.41 \\ 
128 & 257 & 16959 & 32640 & 98.48 & 99.21 \\ 
256 & 513 & 66815 & 130816 & 99.23 & 99.60 \\ 
512 & 1025 & 264959 & 523776 & 99.61 & 99.80 \\ 
1024 & 2049 & 1054719 & 2096128 & 99.80 & 99.90 \\ \hline
\multicolumn{6}{|c|}{{[}1{]}-Proposed design, {[}2{]}- S.Kotiyal et.al \cite{kotiyal2014circuit}, {[}3{]}-R.Zhou et.al \cite{zhou2011transistor} } \\ \hline
\end{tabular}
\end{table}
\begin{table}[]
\centering
\caption{Garbage Outputs Comparison for NxN Reversible Multiplier}
\label{tabmulcomp2}
\begin{tabular}{|c|c|c|c|}
\hline
$ N $    & \begin{tabular}[c]{@{}c@{}}Garbage outputs\\  in   {[}1{]}\end{tabular} & \begin{tabular}[c]{@{}c@{}}Garbage outputs\\ in   {[}2{]}\end{tabular} & \begin{tabular}[c]{@{}l@{}}\% Imp\\  over \\ {[}1{]} \& {[}2{]}
\end{tabular} \\ \hline
4    & 22                                                                      & 36                                                                     & \multirow{9}{*}{100\%}                                                       \\ 
8    & 81                                                                      & 168                                                                    &                                                                            \\ 
16   & 300                                                                     & 720                                                                    &                                                                            \\ 
32   & 1131                                                                    & 2976                                                                   &                                                                            \\ 
64   & 4346                                                                    & 12096                                                                  &                                                                            \\ 
128  & 16953                                                                   & 48768                                                                  &                                                                            \\ 
256  & 66808                                                                   & 195840                                                                 &                                                                            \\ 
512  & 264951                                                                  & 784896                                                                 &                                                                            \\ 
1024 & 1054719                                                                 & 3142656                                                                &                                                                            \\ \hline
\multicolumn{4}{|c|}{{[}1{]}-S. Kotiyal et.al\cite{kotiyal2014circuit}, {[}2{]}-R. Zhou et.al\cite{zhou2011transistor}}                                                                                                                                                    \\ \hline
\end{tabular}
\end{table}
Here, we compare our work with Karatsuba multiplier design presented in \cite{portugal2006reversible}. The comparison for gate count, ancilla inputs, and delay are shown in Table \ref{tabcompkarat}. The design of Karatsuba (1) shown in Table \ref{tabcompkarat} follows Bennet's first scheme. An extra register is used to store the result and the circuit is run backward. The parallel recursive calls are made to reduce the time complexity. The design of Karatsuba (2) also uses parallel recursive call, but the design does not follow Bennet's first scheme, instead  it follows a recursive garbage disposal scheme. The multiplication is computed parallel to garbage disposal. But the trade-off is that the number of gates increases due to the different design blocks adapted in the garbage disposal design process. The design of Karatsuba (3) follows Bennet's first scheme for garbage disposal. The only difference with respect to the design of Karatsuba (1) is that the recursive calls are sequential rather than parallel. Karatsuba (4) is designed using the recursive scheme similar to the one adapted in Karatsuba (2), but recursive calls are sequential. For the designs presented in \cite{portugal2006reversible}, we considered the minimum bound on ancilla inputs calculation. It is observed from Table \ref{tabcompkarat} that with the slight increase in the delay and gate count, the proposed design has improved the ancilla inputs compared to all the Karatsuba designs.
\begin{table}[H]
\centering
\caption{Comparison with Karatsuba Recursive Multiplier}
\label{tabcompkarat}
\begin{tabular}{llll}
\hline
{Designs}        &  {Gate count}     &  {Ancilla inputs}       &  {Delay}      \\
\hline
K(1)                 & $ O $($ n^{log_{2}3} $)             & 6$ n $                  & $ O $($ n $)              \\
K(2)                 & $ O $($ n^{log_{2}6} $)             & 4$ n $                  & $ O $($ n^{log_{2}6} $)        \\
K(3)                 & $ O $($ n^{log_{2}3} $)             & $ 5n+n/2+1 $            & $ O $($ n^{log_{2}3} $)        \\
K(4)                 & $ O $($ n^{log_{2}6} $)             &$  3n+n/2 $              &$  O $($ n^{log_{2}6} $)        \\
Proposed             & $ O $( $ n^{2} $)                 & $ 2n + 1 $              & $ O $($ n^{2} $)             \\
\hline
\multicolumn{4}{c}{K(1),K(2),K(3), and K(4) indicates Karatsuba designs}\\
\multicolumn{4}{c}{1,2,3, and 4 proposed in \cite{portugal2006reversible}}\\
\hline
\end{tabular}
\end{table}

\section{Conclusion}
In this work, we have proposed  ADD and Rotate based on an $ N $x$ N $ reversible multiplier design. We presented the general behavioral model of the design. The proposed multiplier is compared with the relevant existing reversible multiplier designs in the literature. We presented the generalized equations for the performance parameters of the proposed reversible multiplier. It is observed from comparison results that our work outperforms the other designs in terms of the ancilla inputs and zero garbage outputs. The proposed design can be integrated in a larger data path subsystem designs where the garbage outputs and ancilla inputs reductions are the major concerns.


%



\ifCLASSOPTIONcaptionsoff
  \newpage
\fi

\bibliographystyle{IEEEtran}

\bibliography{references-mul-square}




\end{document}